\documentclass[doublecol,figures]{epl2}
\usepackage{amssymb}
\usepackage{amsmath}
\usepackage{dcolumn}
\usepackage{bm}
\usepackage{textcomp} 
\usepackage{units} 
\usepackage[latin1]{inputenc}

\newcommand{\degrees}{\ensuremath{^\circ}\,}

\title{Magnetic coupling in highly-ordered \chem{NiO}/\chem{Fe3O4}(110): \\ Ultrasharp magnetic interfaces vs. long-range magnetoelastic interactions}
\shorttitle{Magnetic coupling in highly-ordered \chem{NiO}/\chem{Fe3O4}(110)}

\author{I. P. Krug\inst{1} \and F. U. Hillebrecht\inst{1} \and H. Gomonaj\inst{2} \and M. Haverkort\inst{2} \and A. Tanaka\inst{3} \and L. H. Tjeng\inst{4} \and C. M. Schneider\inst{1}}
\shortauthor{I. P. Krug \etal}

\institute{
\inst{1} Inst.~f.~Festk\"orperforschung IFF-9, Forschungszentrum J\"ulich GmbH -52425 J\"ulich, Germany\\
\inst{2} Bogolyubov Institute for Theoretical Physics NAS of Ukraine - st.~Metrologichna, 14-b, 03143, Kiev, Ukraine\\
\inst{3} Department of Quantum Matter, ADSM, Hiroshima University - Higashi-Hiroshima, 739-8530, Japan\\
\inst{4} Physikalisches Institut II, Universit\"at zu K\"oln - 50937 K\"oln, Germany
}


\abstract{
  We present a laterally resolved x-ray magnetic dichroism study of
  the magnetic proximity effect in a highly ordered oxide system,
  i.e.~NiO films on Fe$_3$O$_4$(110). We found that the magnetic
  interface shows an ultrasharp electronic, magnetic and structural
  transition from the ferrimagnet to the antiferromagnet. The
  monolayer which forms the interface reconstructs to NiFe$_2$O$_4$
  and exhibits an enhanced Fe and Ni orbital moment, possibly caused
  by bonding anisotropy or electronic interaction between Fe and Ni
  cations. The absence of spin-flop coupling for this crystallographic orientation can be explained by a structurally uncompensated interface and additional magnetoelastic effects.}

\begin{document}

\maketitle

\section{Introduction}
\label{sec:intro}
Many of today`s spintronics devices make extensive use of magnetic
coupling phenomena, in particular, through non-magnetic interlayers or between antiferromagnetic (AF) and ferr(o/i)magnetic (henceforth labeled F(I)M) constituents. The latter coupling is well-known to give rise to the so-called exchange anisotropy or ``exchange bias'' \cite{NS99}. In spite of
huge scientific efforts in this field, the relation between the
exchange biasing phenomenon and the microscopic spin configurations in
both constituents and across the interface is still a matter of
debate. In addition, a detailed experimental insight into these
magnetic proximity effects is often compromised by the imperfection of
the interface and the unknown role of defects. A key factor in discriminating different magnetic coupling mechanisms and elucidating their physical origin is the crystalline and chemical perfection of the sample. In order to unequivocally address the details of the spin-dependent coupling mechanisms, highly ordered systems with well-defined interface roughness and good crystallinity are mandatory. This gives also a chance to make better contact to the various theoretical models. 

Mean field calculations have shown that in case of an ideal crystalline system
with only nearest-neighbor interactions, the interaction zone can be
extremely narrow, in the order of a few monolayers on either side of
the interface \cite{JDK05}. Even an atomically sharp transition is
possible, as has been found experimentally for highly-ordered MnPt/Fe
systems \cite{FBPH06}. The magnetic structure of this \emph{planar domain wall} plays a key role for exchange bias, since it determines
whether or not Zeeman energy can be stored reversibly, if an external field
is applied (so-called exchange spring) \cite{SLAO04}. An important source of the magnetic proximity effect is the variation of the size
and relative orientation of the spin-and orbital moments in the
vicinity of the interface, caused by electronic interaction of
the two layers in contact \cite{QH03,TVTS03}. Even violations of
Hund's third rule were predicted, i.e. the mutual orientation of spin-
and orbital moment is not dominated by the filling of the bands
carrying the magnetic moment, but rather by ligand field and
hybridization effects \cite{QH03,TVTS03}.  Another remarkable feature
in this context is the prediction of a reversal of the uncompensated
magnetization in the antiferromagnet from one interfacial layer to the
next, which is induced by the interplay of the unidirectional
interface anisotropy and the antiparallel coupling of neighboring
atoms within the antiferromagnet \cite{Fin04,JDK05}.

In this contribution, we report the observation of a pronounced proximity
effect in thin NiO films grown on ferrimagnetic Fe$_3$O$_4$(110)
single crystals. We find strong evidence for an atomically sharp
electronic, magnetic and structural transition with a collinear
coupling between the AF and FIM. The details of the coupling show no
evidence of a sign reversal in the uncompensated magnetization for
successive interfacial monolayers in the antiferromagnet. Already the
second monolayer appears to be fully spin-compensated, with vanishing spin and orbital moment. 

\section{Experimental details}
\label{sec:exp_details}

Our choice of NiO/Fe$_3$O$_4$(110) represents an almost ideal model
system, since the small lattice mismatch (0.5\%) results in
pseudomorphic growth and sharp interfaces
\cite{RCSL97,WGAH04}. Contrary to metal/oxide interfaces, which are
often diffuse due to chemical interface reactions
\cite{ROSN01,TMHK06}, the density of defect spins in a purely oxidic
system is generally thought to be very low. In a highly-ordered
crystalline system we can thus expect to find well-defined magnetic
interfaces. We use soft x-ray photoelectron emission microscopy (PEEM)
to arrive at an element-sensitive and spatially-resolved vectorial
magnetometry of the individual FIM and AF constituents by exploiting
circular (XMCD) and linear magnetic dichroism (XMLD), respectively.


The measurements were carried out at the BESSY UE-56/1 SGM beamline
using an Elmitec PEEM III (resolution $<$ \unit[100]{nm}), equipped with
an \textit{in-situ} preparation facility. The incidence angle to the
surface was fixed to \unit[16]{\degrees}, with a degree of circular
(linear) x-ray polarization of typically $>\unit[90]{\%}$. The photon
energy resolution was set to \unit[$<0.2$]{eV}.
\begin{figure}[htbp]
\centering
  \onefigure[width=\linewidth]{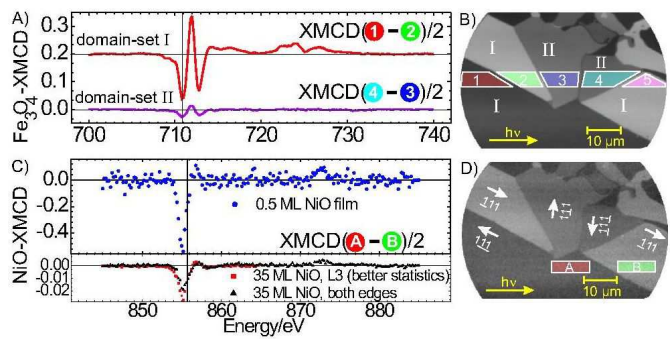}
  \caption{\label{fig:FeONiOXMCD} (color online) (A) Fe$_3$O$_4$-XMCD
    microspectra. Vertical line: Energy position for the ratio image. (B)
    Fe$_3$O$_4$: Ratio image $\sigma^+/\sigma^-$. The numbers 1-5 represent
    areas of interest (AOI) for the microspectra. Latin numbers were assigned to classify domains by their easy axes: $[\underline{1}11]\rightarrow(I)$ and $[1\underline{1}1]\rightarrow(II)$. (C) NiO: XMCD microspectra for a 0.5
    and \unit[35]{ML} NiO film. Vertical line: Energy position for
    ratio image. (D) NiO: XMCD ratio image and magnetization map
    derived from the spectra (white arrows: Fe$_3$O$_4$ and NiO net
    magnetization).}
\end{figure}
Our substrates were synthetic magnetite single crystals,
sputter-cleaned with \unit[1] {keV} Ar ions and subsequently annealed
in $\unit[10^{-6}]{mbar}$ O$_2$ at \unit[1100]{K} for several hours.
After verifying the Fe$_3$O$_4$ phase by x-ray absorption spectroscopy
(XAS) and XMCD, NiO was deposited by molecular beam epitaxy under normal incidence in $\unit[10^{-6}]{mbar}$ O$_2$ background pressure at \unit[300]{K} ($p_{\rm base} <
\unit[2\cdot10^{-9}]{mbar}$).  The low deposition temperature was
chosen on purpose to avoid thermal intermixing at the interface. Wang \emph{et al.} have shown that in such films the electronic transition from
Fe$_3$O$_4$ to NiO at the interface is nearly atomically sharp
\cite{WGAH04}.

\section{Closure domain structure of Fe$_\mathbf{3}$O$_\mathbf{4}$(110)}
\label{sec:magn_domains}
Fig.~\ref{fig:FeONiOXMCD} shows a typical domain pattern of the
(110)-oriented Fe$_3$O$_4$ substrate measured by XMCD at the
Fe L$_3$-edge. Since two of the easy axes of magnetite are coplanar with the
$(110)$-interface, the resulting surface closure-domains will consist
of two sets of \unit[180]{\textdegree}-domains, each belonging to an
easy axis \cite{OXD95}.  As can be seen in fig.~\ref{fig:FeONiOXMCD}B, the
stronger contrast levels (black, white) belong to the
$[\underline{1}11]$-direction and are labeled \emph{Set I}. The
intermediate gray levels belong to magnetization directions almost
perpendicular to the (horizontal) light incidence direction, and can
thus be attributed to the $[1\underline{1}1]$-direction (\emph{Set
  II}). A quantitative comparison of the XMCD contrast in both sets
from the spectra in (A) yields an angle of \unit[$109\pm
1$]{\textdegree} between both easy axes, which is close to the
theoretically expected value of twice the ``magic'' angle \unit[54.73]{\textdegree}. As
verified from the XMCD contrast and Laue diffraction measurements, the
straight domain boundaries run along $\langle111\rangle$-type
directions as well. Thus, we conclude that the magnetization inside the
domains indeed points along the in-plane easy axes.  We also performed
XMCD-measurements at the Ni L$_3$-edge, which yield information about the
uncompensated magnetization in the AF. The contrast levels in
fig.~\ref{fig:FeONiOXMCD}D are identical to fig.~\ref{fig:FeONiOXMCD}B, i. e. the Ni
moment is parallel to the Fe moment. We will discuss this in more
detail in section~\ref{sec:proximity}.

\section{Spin axis orientation of the antiferromagnet}
\label{sec:AF_spinax}
To see how the antiferromagnetic part of the film couples to the Fe$_3$O$_4$ surface, we determined the orientation of the spin axis in NiO exploiting the
linear dichroism at the L$_2$ edge. The absence of a shift of the NiO
L$_3$ peak position between p- and s- polarization clearly proves the
crystal field dichroism to be negligible \cite{HCHA04}. Temperature
dependent XMLD measurements performed after our experiment verified a
lowered blocking temperature around \unit[480]{K} due to finite size
effects \cite{ATVH98}. Thus, we conclude that the observed contrast is
of purely magnetic origin.

\begin{figure}[htbp]
\centering
  \onefigure[width=0.8\linewidth]{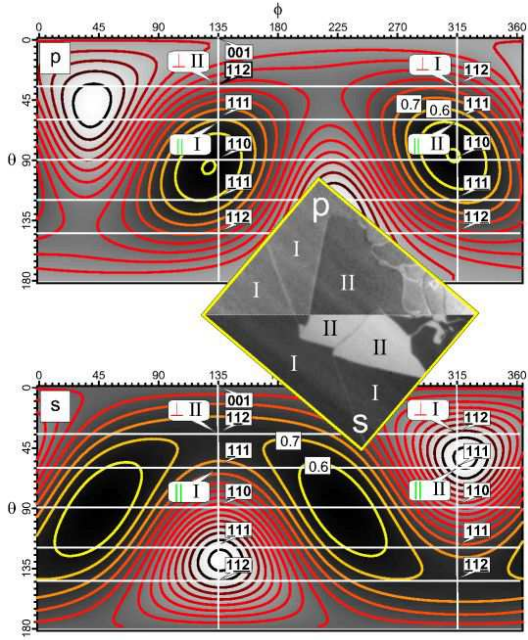}
  \caption{\label{fig:NiOXMLD}(color online) \unit[35]{ML} NiO on
    Fe$_3$O$_4$-(110). The contour plots show the calculated L$_2$
    asymmetry for every possible direction of the spin (angles
    $\theta,\phi$). Center: compilation-image of PEEM p-contrast
    (upper half) and s-contrast (lower half). The two domain sets
    (gray levels) are named I and II. In the contour plots, the
    corresponding crystallographic directions are labelled. Collinear
    coupling: $[\underline{1}11]$ for set I and $[1\underline{1}1]$
    for set II. Conversely for spin-flop coupling, the assignment is
    $[1\underline{1}2]$ for set I and $[\underline{1}12]$ for set
    II. Only the collinear case matches the experimentally determined contrast, with
    set II being brighter in s- and slightly darker in p-geometry. Spin-flop
    coupling would produce the reverse contrast and can thus be
    excluded.}
\end{figure}

The experimental approach that we have chosen is similar to the one in
refs. \cite{ATVH98,HOWB01,OSNA01}, where the ratio of the
multiplet-split NiO L$_2$ peaks was evaluated. However, since with our epitaxial NiO films we have
a single-crystalline material with O$_h$ symmetry, the simple XMLD
relation used in these previous analyses breaks down. It is valid only for
orientation-averaged measurements, where any effects of the site
symmetry will drop out. It has been shown recently, that in oriented
single-crystalline materials, the XMLD is \emph{anisotropic},
i.e. depends on the spin orientation with respect to the crystal
lattice \cite{CNHW06,AVCS06}. We employ a model that is able to
predict the XMLD angular variation for arbitrary spin orientations,
allowing for a quantitative vectometry based on two fundamental
spectra derived from atomic multiplet calculations.

The tiled center image (fig.~\ref{fig:NiOXMLD}) represents the local
L$_2$ ratio in p- and s-contrast, which is related to both the
orientation of the linear polarization $\mathbf{E}$ of the photon field and the
orientation of the spin $\mathbf{S}$ with respect to the cubic crystal
axes\cite{AVCS06}. Our calculations (see contour plots in
fig.~\ref{fig:NiOXMLD}) show best agreement with the experiment for a
collinear coupling, i.e. the spin-axis of the AF is in-plane and oriented along
$\left[\underline{1}11\right]$ or $\left[1\underline{1}1\right]$. At a 
first glance this seems to disagree with theory \cite{Koo97}, since
the NiO(110) interface should be fully compensated, if the bulk AF
structure prevails at the interface. Our analyis will show that the
Fe$_3$O$_4$/NiO \emph{interface} is \emph{not} compensated in the
sense of Koon's theory, so the precondition for spin-flop coupling is actually
not fulfilled.

\begin{figure}[htbp]
\centering
  \onefigure[width=0.8\linewidth]{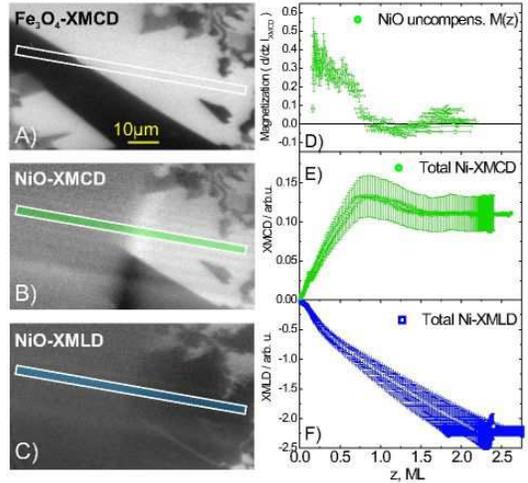}
  \caption{\label{fig:Wedg_lineprof}(color online) NiO wedge on
    Fe$_3$O$_4$.  Left column: PEEM images with line profile position
    indicated. (A) Fe XMCD (profile not shown), (B) Ni XMCD ratio
    image, (C) NiO L$_2$ ratio image for s-polarized light.  Right
    column: Thickness-dependent line profile data : D) Depth-profile of
    $M_{AF}(z)$ (derivative of total XMCD signal). E) Total Ni-XMCD
    signal vs. thickness. F) Total Ni-XMLD vs. thickness.}
\end{figure}

\section{Magnetic structure of the interface region}
\label{sec:proximity}

We used XMCD at the Ni $L$-edges to selectively study the
magnetization in the adlayer (fig.~\ref{fig:FeONiOXMCD}C and
\ref{fig:FeONiOXMCD}D). In order to prove that the NiO magnetization induced by the
contact to the FIM is confined to the interface region, we compare two
cases: a fractional \unit[0.5]{ML} NiO coverage and a thick
(\unit[35]{ML}) film. The resulting spectra are shown in
fig.~\ref{fig:FeONiOXMCD}C. For \unit[0.5]{ML} we get a maximum
dichroic contrast of \unit[54]{\%} in the white line. Both the
magnitude and the spectral shape of the dichroism closely match the
results of v.~d.~Laan (\unit[53]{\%}) for NiFe$_2$O$_4$
\cite{LHPD99}. This means that the Ni moments are parallel to
Fe$_3$O$_4$ and located at sites with octahedral oxygen
coordination. From the spectra alone, NiFe$_2$O$_4$ and NiO cannot be
distinguished, since Ni has the same oxygen coordination in both cases. As
pointed out by Wang et al., the NiO/Fe$_3$O$_4$ interface may in fact
reconstruct to NiFe$_2$O$_4$, which is isostructural to Fe$_3$O$_4$
\cite{WGAH04}. In the absence of thermal intermixing, this phase
should be confined to one interfacial layer only. In contrast to the strong
dichroism at \unit[0.5]{ML} coverage we found a strongly reduced contrast of
only \unit[2.7]{\%} in the \unit[35]{ML}-film. Comparison of the
intensity-normalized XMCD in both cases yields a rough estimate of
\unit[2.6]{\AA}(\unit[1.7] {ML}) contributing to the XMCD signal. This
low value thus indicates a good interface quality and has already been
predicted for bilayers with ideal crystalline structure \cite{JDK05}.

\begin{figure}[htbp]
  \onefigure[width=0.8\linewidth]{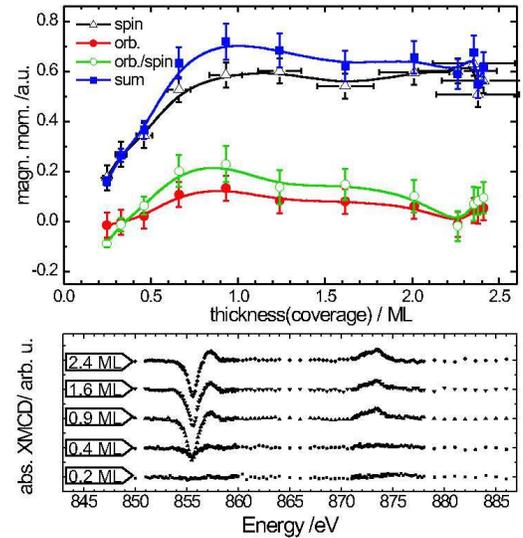}
  \caption{\label{fig:wedg_microspec}(color online): Local NiO XMCD
    microspectra and sum rule analysis. The upper panel shows the
    thickness-dependent spin- (open triangles) and orbital moments
    (circles), the ratio $m_{\rm orb.}/m_{\rm spin}$ (open circles)
    and the sum $m_{\rm orb.}+m_{\rm spin}$. (squares). The spin
    moment increases up to one monolayer and then stays constant for
    higher thicknesses, while the orbital moment shows a pronounced
    maximum near \unit[1]{ML} and then decreases again. The origin of
    the minimum at \unit[2.3]{ML} is still unclear. It could be caused
    by the completion of a second monolayer, but it could also be an artifact,
    since only a single data point is affected. Data points at higher
    thicknesses show a slightly increasing trend, but within the error margin their values are
    essentially the same as above \unit[1]{ML}. Assuming an artifact,
    then $m_{\rm orb.}$ is constant for thicknesses greater than
    \unit[1.5]{ML}. The sum of orbital and spin moment closely
    resembles the curve already gained from the XMCD line profile, so
    the pronounced maximum near \unit[1]{ML} is definitely caused by
    the orbital moment.}
\end{figure}

In order to determine the magnetic microstructure of the narrow proximity zone near the interface, a NiO stepped wedge was grown onto
Fe$_3$O$_4$(110), with a step height of \unit[2.5]{ML} (\unit[3.5]
{\AA}) and $\sim$\unit[20]{\textmu m} wide step slopes. Since the
interesting effects appear within the first few monolayers, we
concentrate on the first step slope. In order to monitor the thickness
dependent changes in magnetic structure, we rely on the analysis of
both image line profiles and microspectra. The line profiles were taken
from PEEM parameter images, usually division images of two helicities
(XMCD) or energies (XMLD), and provide a quick and convenient way to
unravel the thickness-dependent changes in magnetic structure (see
fig.~\ref{fig:Wedg_lineprof}A-C). The XMCD microspectra
(fig.~\ref{fig:wedg_microspec}), in contrast, allow for a more
detailed analysis with separation of spin- and orbital
contributions. In order to reduce intensity fluctuations, each XMCD spectrum was computed from a pair of areas of interest (AOI) (rectangles in fig.~\ref{fig:wedg_microspec}) belonging to a pair of 180\textdegree-domains. The
corresponding thickness calibration was done using the isotropic L$_3$
intensity, which at low thicknesses is proportional to the
coverage. The latter was determined by quartz-balance evaporation
rate measurements prior to and after deposition.  In both the Ni-XMCD
line profile (fig.~\ref{fig:Wedg_lineprof}E ) and the total magnetic
moment (blue squares in fig.~\ref{fig:wedg_microspec}), the same
behaviour can be observed: The total moment increases up to a maximum, which is reached at a coverage near \unit[1]{ML}. As soon as the
second NiO monolayer starts to nucleate, the total moment decreases again and stays approximately constant above a thickness of $\approx\unit[1.5]{ML}$. In contrast to the XMCD, the XMLD signal increases (decreasing gray
level) with a slight bend around \unit[0.3]{ML} and approximately
linear behaviour until it levels off at the step terrace.

The separation of $m_{\rm orb}$ and $m_{\rm spin}$ by a sum rule analysis shows that the extremum observed at one monolayer coverageis caused by the
\emph{orbital moment}, while the total spin moment increases until the
first monolayer is completed. It then stays constant if further material is
deposited on top (see fig.~\ref{fig:wedg_microspec}). This means that
the NiO layer assumes its compensated antiferromagnetic structure
already in the second monolayer, i.e. the uncompensated moments reside
directly at the interface. Consequently, there is no planar domain
wall forming in the ground state of the system, and the magnetic
transition from ferri- to antiferromagnet is atomically sharp -- in
accordance with the findings from the electronic and structural transitions. This result is reasonable, since the magnetic interaction length is in the order of the lattice constant in poorly conducting correlated materials such as transition
metal oxides.

\begin{figure}[htbp]
\centering
  \onefigure[width=0.95\linewidth]{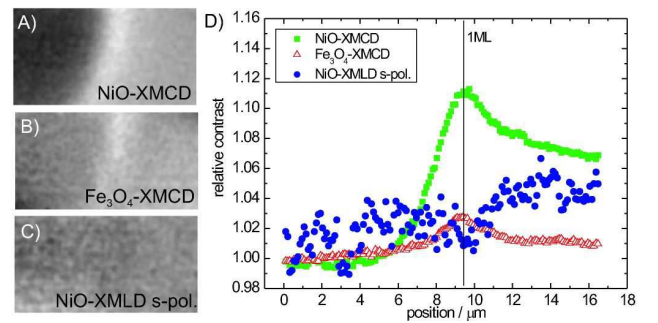}
  \caption{\label{fig:fe_prox}(color online) Enlargement of the slope
    area. Clearly, in all signals, an enhancement around one monolayer
    coverage can be observed, although in the Fe-XMCD and the Ni-XMLD
    the effect is only a few percent.}
\end{figure}

In fig.~\ref{fig:fe_prox}, the slope region around one monolayer
thickness is shown in more detail, now also including the
Fe$_3$O$_4$-XMCD contrast profile. In all contrast patterns there is a clear
extremum strictly confined to the thickness range around one
monolayer. The peak in the Ni-XMCD as well as the coinciding dip in the
Ni-XM\emph{L}D indicate that an extremal value of the total
magnetic moment magnitude $|\langle\boldsymbol{\mu}\rangle|$ and
consequently also $|\langle\boldsymbol{\mu}^2\rangle|$ must occur. The
Fe$_3$O$_4$ XMCD contrast is enhanced by $\unit[3]{\%}$, which
corresponds -- depending of the probing depth ( $\lambda_e\approx\unit[10]{\mbox{\AA}}$ \cite{GGLS06} or $\lambda\approx\unit[50]{\mbox{\AA}}$ \cite{GGS00}) -- to an
enhancement of the total moment at the interface in the range of
\unit[120-200]{\%}.  This increase could be caused by the Fe orbital
moment -- similar to Ni. On the other hand, it could also be a consequence of an electronic interaction between the Ni and Fe sites, for example, if the
interfacial monolayer reconstructs to NiFe$_2$O$_4$.

Note that Lueders \textit{et al.} have found a considerable
enhancement of the total magnetic moment up to \unit[250]{\%} in
ultrathin NiFe$_2$O$_4$ films \cite{LBBC05}. If this effect is an
intrinsic property of low-dimensional NFO, it could support the
hypothesis of a NiFe$_2$O$_4$ interface layer. Provided that only one
monolayer shows and enhanced Fe-XMCD signal, a \unit[3]{\%} contrast enhancement with a probing depth of \unit[50]{\AA} would correspond to an
enhancement as large as \unit[200]{\%}, which comes close to the
results in Ref.~\cite{LBBC05}.

\section{Discussion of the results}
\label{sec:discuss}
Since a bulk-truncated NiO(110) surface is atomically compensated, the occurence of collinear exchange coupling in our system is astonishing and at first glance contradicts the findings of Koon \cite{Koo97}. In the following, we will discuss possible reasons for the discrepancy.
First, we have to consider the interface between the two materials Fe$_3$O$_4$ and NiO in more detail. If we compare the two structures, it becomes apparent that although magnetite has almost twice the crystallographic lattice constant of NiO, the \emph{magnetic} unit cells of both materials match at the interface. This means that the two sublattices of NiO can experience different magnetic environments at the interface, leading to nondegenerate interface exchange constants $J_1\neq J_2$. If the imbalance between these two coupling configurations is large enough, a spin-flop state is no longer stabilized and collinear coupling can occur. This is especially true, if we omit the somewhat idealized picture of bulk-truncated surfaces. Assuming for the moment that the interface layer reconstructs to NiFe$_2$O$_4$, Ni-cations will be located at octahedral positions only and the imbalance between $J_1$ and $J_2$ is enhanced. 

These considerations are able to describe our findings for the (011)-interface. However, they have important implications for the coupling between the two materials in general, because this imbalance situation holds also for other surface orientations. Therefore, one would in fact expect collinear coupling for arbitrary interface orientations. In another study, however, we found spin-flop coupling for Fe$_3$O$_4$(001)/NiO \cite{TBP}. In general, the (001) interface is atomically compensated and spin-flop coupling should therefore occur quite naturally, but the lifting of the NiO sublattice degeneracy by the inverse spinel structure of the magnetite discussed above should suppress this type of coupling. Thus, there must still be another mechanism at work, which introduces a dependence on the interface orientation. As will be discussed in the following, we believe that magnetoelastic effects -- which have been neglected so far in most studies -- play an important role in the magnetic coupling process.

Magnetoelastic effects, e.g. introduced by a lattice misfit are well-known to influence the magnetic behavior in ferromagnetic thin films. Similar effects must also be expected for antiferromagnets. Krishnakumar \textit{et al.} \cite{KLGV07} observed a thickness-dependent change in the magnetic anisotropy in Ag(001)/NiO, which they explained by strain relaxation, thereby involving magnetoelastic effects. Their hypothesis is supported by the finding that the presence of a MgO capping layer, which introduces additional strain, considerably weakened the thickness-dependent change in anisotropy. It should be noted in this context that Finazzi \textit{et al.} have observed a similar transition from collinear to spin-flop coupling in Fe(001)/NiO at a critical AF thickness, but explained the effect by defects \cite{FBBG06}. However, also in their case magnetoelastic effects might be of importance, since the lattice mismatch between the R45 NiO epitaxial growth and Fe is considerable (NiO is  compressed in-plane by $\approx\unit[-3]{\mbox{\%}}$). In our case of the $(110)$-surface, the lattice mismatch of NiO is considerably lower (in-plane tensile \unit[0.5]{\%}). Nevertheless we propose that magnetoelastic effects determine the coupling, and we will support this idea by qualitative arguments.
\begin{figure}[htbp]
\centering
\onefigure[width=\linewidth]{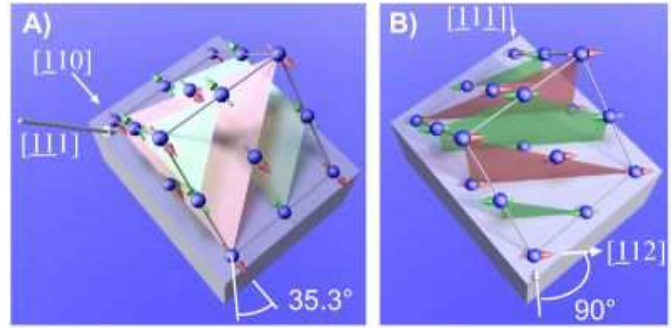}
\caption{\label{fig:NiO_stacking}(color online) Two types of strain-induced AF stacking in Fe$_3$O$_4$(110)/NiO. A) the tensile in-plane epitaxial strain of NiO causes out-of-plane compression and stacking (for example along $[\underline{11}1]$ as indicated in the figure). The intersection of the easy planes with the $(110)$-interface is the $[\underline{1}10]$-direction. B) A hypothetical in-plane compression along one of the magnetite easy axes causes the spins of NiO to be perpendicular to magnetite, along the intersection of the easy planes and the interface. This case is never realized since the in-plane epitaxial strain is tensile, and moreover the magnetostriction in magnetite is positive along $\langle111\rangle$.}
\end{figure}

A single-domain NiO crystal shows a contraction by \unit[-0.15]{\%} along its $\langle111\rangle$-type stacking direction. Conversely, in a strained NiO layer those $\langle111\rangle$ directions, which are compressed most by the epitaxial strain, become favorable. For a Fe$_3$O$_4$(110)/NiO interface this means that out-of plane stacking should be favored over in-plane stacking (as in fig.~\ref{fig:NiO_stacking}A ) . If in-plane stacking were to occur (fig.~\ref{fig:NiO_stacking}B ), the stacking vectors would be parallel to the in-plane easy directions ($[\pm1\mp11]$)in magnetite, rendering the easy planes perpendicular to the interface. The intersecting in-plane easy directions of NiO would then be the $[\mp1\pm12]$-directions, which in case of a stacking parallel to the Fe$_3$O$_4$ magnetization lead to spin-flop coupling (fig.~\ref{fig:NiO_stacking}B ). All other configurations result in collinear coupling. Especially, since the actual stacking preferred by the epitaxial strain is out-of-plane, the easy directions of NiO are along $[\underline{1}10]$, and thus closer to the magnetite easy axes (see fig.~\ref{fig:NiO_stacking}A ). In this situation, collinear coupling is favoured, but the interfacial anisotropy will contain a uniaxial contribution along $[\underline{1}10]$. In our as-grown samples, this contribution is not measurable, but becomes apparent, if we anneal the samples \cite{TBP2}. Finalizing our discussion, one could say that spin-flop coupling is "exotic" in our system, since it requires the right strain situation and/or a compensated interface. Both aspects are not realized and collinear coupling results naturally.

As a consequence of the above discussion, the coupling at the Fe$_3$O$_4$(110)/NiO interface should be seen as a delicate balance between at least two mechanisms: (i) a competition between exchange coupling contributions from the FIM and AFM sublattices and (ii) magnetoelastic interactions. As both mechanisms can independently lead to either collinear or spin-flop coupling, depending on the relative strength of the interactions, an a priori prediction of the coupling type for a given system is not trivial. This situation should also be encountered in many other exchange-bias systems, in particular, if a sizable lattice mismatch exists.

\section{Summary and Conclusions}
\label{sec:sumconcl}
In summary, we found an enhanced Fe and Ni magnetization directly at
the NiO/Fe$_3$O$_4$(110) interface, which has its maximum at a
coverage equivalent to \unit[1]{ML}, implying a reconstruction of the
interfacial monolayer towards the NiFe$_2$O$_4$ structure. As a direct
consequence, the interface is not atomically compensated and collinear
coupling can occur, in agreement with Koon's theory\cite{Koo97}. Another explanation for the collinear coupling lies in the epitaxially strained NiO layer, in which out-of-plane AF stacking vectors are preferred due to magnetoelastic effects.
Sum rule analysis indicates that the extremum in the interfacial Fe and Ni total
moments at one monolayer coverage might be caused by an enhanced
orbital moment due to bonding anisotropy or an interaction of Ni and
Fe cations in a NFO reconstructed interface layer. We found no
evidence for a sign-reversal of the uncompensated magnetiztion in the
antiferromagnet as postulated theoretically \cite{Fin04,JDK05}. On the
contrary, it appears that the electroncic, magnetic and structural
transition at the interface is atomically sharp and that already the
second monolayer in the AF is fully compensated. 
\acknowledgments
This work was partially funded by the Sonderforschungsbereich 491. We
thank Y.-X. Su, D. Schrupp for supplying the synthetic magnetite
crystals, M. Schmidt and C. Thomas for the Laue
measurements.

\bibliographystyle{eplbib}
\bibliography{NiOFe3O4}

\begin{thebibliography}{10}
\expandafter\ifx\csname url\endcsname\relax\def\url#1{\texttt{#1}}\fi

\bibitem{NS99}
\Name{Nogues J. \and Schuller I.~K.} \REVIEW{J. Magn. Magn. Mat.
  }{192}{1999}{203}.

\bibitem{JDK05}
\Name{Jensen P.~J., Dreyssé H. \and Kiwi M.} \REVIEW{Eur. J. Phys. B
  }{46}{2005}{541}.

\bibitem{FBPH06}
\Name{Fujii J., Borgatti F., Panaccione G., Hochstrasser M., Maccherozzi F.,
  Rossi G. \and van~der Laan G.} \REVIEW{Phys. Rev. B }{73}{2006}{214444}.

\bibitem{SLAO04}
\Name{Scholl A., Liberati M., Arenholz E., Ohldag H. \and Stohr J.}
  \REVIEW{Phys. Rev. Lett. }{92}{2004}{247201}.

\bibitem{QH03}
\Name{Qian X. \and Hubner W.} \REVIEW{Phys. Rev. B }{67}{2003}{184414}.

\bibitem{TVTS03}
\Name{Tyer R., van~der Laan G., Temmerman W.~M., Szotek Z. \and Ebert H.}
  \REVIEW{Phys. Rev. B }{67}{2003}{104409}.

\bibitem{Fin04}
\Name{Finazzi M.} \REVIEW{Phys. Rev. B }{69}{2004}{064405}.

\bibitem{RCSL97}
\Name{Recnik A., Carroll D.~L., Shaw K.~A., Lind D.~M. \and Ruhle M.}
  \REVIEW{J. Mater. Res. }{12}{1997}{2143}.

\bibitem{WGAH04}
\Name{Wang H.~Q., Gao W., Altman E.~I. \and Henrich V.~E.} \REVIEW{J. Vac. Sci.
  Technol. A }{22}{2004}{1675}.

\bibitem{ROSN01}
\Name{Regan T.~J., Ohldag H., Stamm C., Nolting F., Luning J., Stohr J. \and
  White R.~L.} \REVIEW{Phys. Rev. B }{64}{2001}{214422}.

\bibitem{TMHK06}
\Name{Tusche C., Meyerheim H.~L., Hillebrecht F.~U. \and Kirschner J.}
  \REVIEW{Phys. Rev. B }{73}{2006}{125401}.

\bibitem{OXD95}
\Name{Ozdemir O., Xu S. \and Dunlop D.~J.} \REVIEW{J. Geophys. Res.-Sol. Ea.
  }{100}{1995}{2193}.

\bibitem{HCHA04}
\Name{Haverkort M.~W., Csiszar S.~I., Hu Z., Altieri S., Tanaka A., Hsieh
  H.~H., Lin H.~J., Chen C.~T., Hibma T. \and Tjeng L.~H.} \REVIEW{Phys. Rev. B
  - Rapid Comm. }{69}{2004}{020408(R)}.

\bibitem{ATVH98}
\Name{Alders D., Tjeng L.~H., Voogt F.~C., Hibma T., Sawatzky G.~A., Chen
  C.~T., Vogel J., Sacchi M. \and Iacobucci S.} \REVIEW{Phys. Rev. B
  }{57}{1998}{11623}.

\bibitem{HOWB01}
\Name{Hillebrecht F.~U., Ohldag H., Weber N.~B., Bethke C., Mick U., Weiss M.
  \and Bahrdt J.} \REVIEW{Phys. Rev. Lett. }{86}{2001}{3419}.

\bibitem{OSNA01}
\Name{Ohldag H., Scholl A., Nolting F., Anders S., Hillebrecht F.~U. \and Stohr
  J.} \REVIEW{Phys. Rev. Lett. }{86}{2001}{2878}.

\bibitem{CNHW06}
\Name{Czekaj S., Nolting F., Heyderman L.~J., Willmott P.~R. \and van~der Laan
  G.} \REVIEW{Phys. Rev. B - Rapid Comm. }{73}{2006}{020401(R)}.

\bibitem{AVCS06}
\Name{Arenholz E., van~der Laan G., Chopdekar R.~V. \and Suzuki Y.}
  \REVIEW{Phys. Rev. B }{74}{2006}{094407}.

\bibitem{Koo97}
\Name{Koon N.~C.} \REVIEW{Phys. Rev. Lett. }{78}{1997}{4865}.

\bibitem{LHPD99}
\Name{van~der Laan G., Henderson C. M.~B., Pattrick R. A.~D., Dhesi S.~S.,
  Schofield P.~F., Dudzik E. \and Vaughan D.~J.} \REVIEW{Phys. Rev. B
  }{59}{1999}{4314}.

\bibitem{GGLS06}
\Name{Goering E., Gold S., Lafkioti M. \and Schutz G.} \REVIEW{Europhys. Lett.
  }{73}{2006}{97}.

\bibitem{GGS00}
\Name{Gota S., Gautier-Soyer M. \and Sacchi M.} \REVIEW{Phys. Rev. B
  }{62}{2000}{4187}.

\bibitem{LBBC05}
\Name{Luders U., Bibes M., Bobo J.~F., Cantoni M., Bertacco R. \and Fontcuberta
  J.} \REVIEW{Phys. Rev. B }{71}{2005}{134419}.

\bibitem{TBP}
I. Krug \etal, unpublished.

\bibitem{KLGV07}
\Name{Krishnakumar S., Liberati M., Grazioli C., Veronese M., Turchini S.,
  Luches P., S. V. \and Carbone C.} \REVIEW{J. Magn. Magn. Mat.
  }{310}{2007}{8}.

\bibitem{FBBG06}
\Name{Finazzi M., Brambilla A., Biagioni P., Graf J., Gweon G.-H., Scholl A.,
  Lanzara A. \and Duò L.} \REVIEW{Phys. Rev. Lett. }{97}{2006}{097202}.

\bibitem{TBP2}
The influence of annealing on the magnetism in this material system exceeds the
  scope of the present paper and will be treated elsewhere.

\end{thebibliography}

\end{document}